\documentstyle[12pt,aps]{revtex}
\begin{document}
\def\ApJ{{\sl Astrophys. J.}}
\def\la{\mathrel{\mathpalette\fun <}}
\def\ga{\mathrel{\mathpalette\fun >}}
\def\fun#1#2{\lower3.6pt\vbox{\baselineskip0pt\lineskip.9pt
  \ialign{$\mathsurround=0pt#1\hfil##\hfil$\crcr#2\crcr\sim\crcr}}}
\title{\ \\\ \\\ \\\ \\\ \\
The Uncertainties in the Age of Globular Clusters from Their
Helium Abundance and Mass Loss}
\author{X. Shi\\
The University of Chicago, Chicago, IL 60637-1433}
\vskip 3.8cm
\maketitle

\centerline{\bf Abstract}
The age of globular clusters inferred from observations
depends sensitively on assumptions such as the initial helium abundance and the
mass loss rate. A high helium abundance (e.g., $Y\approx$0.28),
as well as an inclusion of helium diffusion and oxygen-enhancement
in stellar models,
can lower the current age estimate for metal-poor globular clusters
from 14$\pm 1.5$ Gyr to about 11$\pm 1$ Gyr, significantly
relaxing the constraints on the Hubble constant,
allowing values as high as 60km/sec/Mpc for
a universe with the critical density and 90km/sec/Mpc for a baryon-only
universe. The uncertainties of a high helium abundance and
an instability strip induced mass loss near the turn-off in globular clusters
are discussed. Ages lower than 10 Gyr are not possible even
with the operation of both of these mechanisms unless the initial helium
abundance in globular clusters is $>0.28$, which can hardly be accomodated
by indirect inferences of helium abundances in globular clusters.
\vskip 0.7in
\noindent --------------------------------------------------------

\noindent{Submitted to \ApJ}
\vfill\eject

\section{Introduction}
The latest estimate of the age of globular clusters (GCs)
is 14.1$\pm 1.5$ Gyr averaging over 24 globular clusters \cite{Sandage93a},
or 15$\pm 1.5$ Gyr for oldest
GCs \cite{Walker1992}. This sets a lower bound on the
age of the universe which
severely constrains $H_0$, the Hubble constant, for a $\Lambda$
(the vacuum energy density)=0 Friedman-Robertson-Walker cosmology.
For example, if the age of the universe is 15 Gyr,
for $\Omega$ (the ratio of the density of the universe to the
critical density)$=1$, the Hubble constant $H_0$
must be less than 45km/sec/Mpc \cite{Sandage93a}; while
for a baryonic universe where $0.01/h^2<\Omega<0.015/h^2$
(with $h=H_0/100$km/sec/Mpc) as constrained
by Big Bang Nucleosynthesis \cite{Walker1991,Copi},
$H_0\la 70$km/sec/Mpc. Currently
dynamic measurements of the density of the universe favor
$\Omega>0.3$ \cite{NusserDekel1993}, which yields $H_0\la 55$ km/sec/Mpc.
Direct measurements of the Hubble constant, however, yield values that
range widely from 40km/sec/Mpc to 100 km/sec/Mpc \cite{Huchra1992}.

A reconcilation of the direct measurements of the Hubble constant
and that inferred from the age of the globular clusters may well
require not only improvements of the distance scale systematics but also
continued investigation of the implicit assumptions in the
inferred age of globular clusters. Otherwise, the conflict may
prove to be an insurmountable problem for $\Omega=1$, $\Lambda=0$ cosmology
models.

It is noted that the estimate of the age depends sensitively on many
uncertain aspects of globular clusters,
such as the actual turn-off luminosity (distance) determination,
the metallicity, the oxygen to iron ratio, helium diffusion in stars,
the initial helium abundance, and
an instability strip induced mass loss inferred from
lithium observations of stars \cite{ShiSchrammDearbornTruran}.
The uncertainties of the turn-off luminosity, the metallicity and
the oxygen to iron ratio have been widely discussed in the
literature \cite{Wheeler89,V92,Bergbusch1992},
and has been included in the current age estimate of $14.1\pm 1.5$
Gyr \cite{Sandage93a}. The helium diffusion in stars is also estimated to
lower the age of GCs by $5\%$--15$\%$
\cite{ProffittMichaud1991,ProffittVandenberg1991,Chaboyer92}. Uncertainties
in the initial helium abundance and the mass loss rate of stars near the
turn-off point have been explored previously
\cite{ShiSchrammDearbornTruran,DearbornSchramm1993}.
But since these systematic uncertainties should not be added in quadrature
like statistical uncertainties, a full exploration including uncertainties
in the helium abundance, mass loss, helium diffusion and oxygen-enhancement
is essential to assess their influence on the age estimate of globular
clusters.

The aim of this paper is to explore the uncertainties in the helium abundance
and mass loss in globular clusters both theoretically and observationally,
and calculate the age estimates for globular clusters under these uncertainties
as well as oxygen-enhancement, using stellar models with helium diffusion.
Section 2 discusses the methods used in dating the age of
globular clusters and their dependence on various parameters. In this section,
we also present our formula that relates the age of globular clusters to
their parameters, based on our stellar models with helium diffusion.
Section 3 discusses uncertainties in the current
inferences of the helium abundance in globular clusters due to uncertainties
in both stellar evolution theory and stellar observations. In particular,
it argues that the current inferences may allow a helium abundance as high as
0.28 in globular clusters,
which is significantly higher than the primordial value of 0.24.
Section 4 provides possible scenarios that
can enhance the helium abundance in globular clusters from the primordial
value to a higher one. Section 5 then discuss a possible
instability strip induced mass
loss near the turn-off region of globular clusters, and its effect on the
luminosity functions (LFs) of globular clusters and their age.
The observed luminosity function of globular clusters
thus constrains the size of the mass loss effect. The effect of a mass loss
combined with a high helium abundance is also investigated. Section 6
summarizes the previous sections and concludes that
by considering the uncertainties in the helium abundance and mass loss in
globular clusters, as well as helium diffusion and oxygen enhancement,
the age of metal-poor clusters can be as low as $11\pm 1$ Gyr.
But ages lower than 10 Gyr for metal-poor clusters
cannot be consistent with observed properties of globular clusters.

\section{Age dating of globular clusters \label{Age dating}}
Two methods are commonly used to estimate the age of a globular cluster
(see review of ref. \cite{Demarque91}).
One is to fit theoretical isochrones (temperature-luminosity
curves for stars of a given age)
to the shape of a globular cluster on the observed color-magnitude diagram,
using a distance modulus determined from fitting either a zero age main
sequence (ZAMS) or a zero age horizontal branch (ZAHB).
In the fitting, it is necessary to transform a temperature-luminosity
diagram of theoretical isochrones into a color-magnitude diagram with an
interstellar reddening correction \cite{Bergbusch1992}.
Besides the uncertainties mentioned in the introduction,
this method is very sensitive to the transformation between the
luminosity-temperature diagram and the color-magnitude diagram,
the fiducial sequence of globular clusters extracted from their scattered
color-magnitude diagram, the subjectivity in fitting isochrones to the
fiducial sequence, and the effectiveness
of a constant mixing length in describing the convection.
It also depends sensitively on the adopted
interstellar reddening if using ZAMS fitting to obtain distances.

The difficulty in improving the fitting method lies mainly in the fact that
the color of both observed GCs and theoretical isochrones on the
color-magnitude diagram is hard to determine precisely.
The color of observed GCs depends on the
interstellar reddening which can be very uncertain;
while the color of isochrones is sensitive to the treatment of convection
in stellar models. Figure 1 shows isochrones with similar parameters but
different mixing lengths (parametrized by $\alpha$, the ratio of
the mixing length to the pressure scale height) in the convection zone.
Both isochrones have a similar turn-off luminosity, but the
color shift due to a different choice of
$\alpha$ is manifest, especially on the upper subgiant and the red giant branch
of isochrones. Although the uncertainty in choosing different
mixing lengths may be partially resolved by calibrations of ZAMS
stars \cite{Vandenberg1988},
it still persists since mixing lengths may change in different
evolution phases, such as the subgiant branch and the red giant branch.

A method that is not sensitive to the color determination is to relate the
age of a GC with the difference in luminosity between the GC's turn-off point
and horizontal branch, regardless of its detailed morphology.
Such a relation has been shown by Iben and Renzini \cite{IbenRenzini1984} to be
\begin{equation}
\log \Bigl({{\rm Age}\over {\rm 1 yr}}\Bigr)
=8.497-1.88(Y-0.24)-1.44(Y_{\rm HB}-Y)-0.088\log Z
+0.41(M^{\rm TO}_{\rm bol}-M^{\rm RR}_{\rm bol}),
\label{Age-LIben}
\end{equation}
where $Y$ is the initial helium abundance of GCs, $Z$ is their metallicity,
$Y_{\rm HB}$ is the helium abundance at the envelope
of the HB stars, $M^{\rm TO}_{\rm bol}$ and $M^{\rm RR}_{\rm bol}$ are the
bolometric magnitudes of the turn-off point and the RR Lyrae stars
respectively.
The coefficients in the equation have been refined over the years, but the
resultant age hasn't changed by more than $25\%$.
For example, a similar relation based on Sandage's calibration
of RR Lyrae stars \cite{Sandage1993b} and the oxygen-enhanced isochrones of
Bergbush and VandenBerg \cite{Bergbusch1992}, assuming $Y\approx 0.24$,
is \cite{Sandage93a}
\begin{equation}
\log \Bigl({{\rm Age}\over {\rm 1 yr}}\Bigr)
=8.773+0.017 [{\rm Fe}/{\rm H}]
+0.39(M^{\rm TO}_{\rm v}-M^{\rm RR}_{\rm v}),
\label{Age-LSandage}
\end{equation}
where $M_{\rm v}$ denotes the visual magnitudes. Equation (\ref{Age-LSandage})
tends to yield an age that is $\sim 20\%$ lower than eq.~(\ref{Age-LIben})
for metal-poor clusters.
The method of relating the age of GCs and their turn-off luminosities
is independent of the interstellar reddening, the subjectivity of fitting
isochrones to the observed GC sequence and the mixing length adopted in
stellar modeling. It still suffers from the uncertainties mentioned
in the introduction.

In order to test the sensitivity of GC ages to the helium abundance
for stellar models with helium diffusion, we constructed a series of
models between 0.60$M_\odot$ and 0.90$M_\odot$, with an interval
of 0.05$M_\odot$. Their metallicities are chosen to be Z=0.0002
($10^{-2}Z_\odot$), 0.0006 and 0.002.
The helium abundance of the models is set to be 0.24 and 0.28.
We adopted a mixing length of $\alpha=1.69$. All models
were started as pre-main sequence objects on the Hayashi track, and
evolved to helium flush, or 25 Gyr old (which is enough for GCs around
15 Gyr old), whichever comes earlier.
We used Livermore opacity tables for temperatures above 6000K \cite{Rogers},
and Los Alamos opacities below 6000K \cite{Heubner}.
The code we used is based on Dearborn's stellar code \cite{Dearborncode},
with an inclusion of helium diffusion using the prescription of Bahcall and
Loeb
\cite{BahcallLoeb}. From these models, we found that
\begin{equation}
\log \Bigl({{\rm Age}\over {\rm 1 yr}}\Bigr)=
8.11-0.12\log Z +0.40 M_{\rm bol}^{\rm TO}-0.80(Y-0.24).
\label{Age-TO}
\end{equation}
Our models agree fairly well with previous helium diffusion
calculations \cite{ProffittMichaud1991,ProffittVandenberg1991,Chaboyer92}.

Since the helium diffusion affects little the evolution of HB stars due to
their short life span, we can just take the result of HB calculations without
including helium diffusion \cite{Sweigart,Buzzoni1983}:
\begin{equation}
M_{\rm RR}{\rm (bol)}=0.943-3.5(Y_{\rm HB}-0.30)+0.183\log Z.
\label{HBLum}
\end{equation}
Combining eq.~(\ref{Age-TO}) and (\ref{HBLum}), we get
\begin{equation}
\log \Bigl({{\rm Age}\over {\rm 1 yr}}\Bigr)
=8.57-0.047\log Z-2.20(Y-0.24)-1.40(Y_{\rm HB}-Y)
+0.40(M^{\rm TO}_{\rm bol}-M^{\rm RR}_{\rm bol}).
\label{Age-L}
\end{equation}
This result compared with non-diffusion calculations based on the
same code \cite{ShiSchrammDearbornTruran} lowers the age
of metal-poor clusters by about 8\%. It yields ages that
are generally 15\% to 25\% lower than eq.~(\ref{Age-LIben}), depending on $Z$.

We couldn't obtain oxygen-enhanced models with Livermore opacity tables.
But it has been shown that for Los Alamos opacity tables,
oxygen-enhanced models with a metallicity $Z$ (deduced from [Fe/H] only)
can be approximated by standard solar-scaled abundance
models with a metallicity $Z_{\rm eff}$, provided that \cite{Salaris}
\begin{equation}
Z_{\rm eff}=(0.64f+0.36)Z,
\label{Oxygen}
\end{equation}
where $f$ is the enhancement factor of oxygen.
If this also holds true for Livermore Opacity Tables, oxygen-enhanced models
will yield the same equation as eq.~(\ref{Age-L}), as long as $Z_{\rm eff}$ is
substituted for $Z$. Commonly $f$ is taken to be 3 to 5 in metal poor
clusters \cite{V92}, therefore $Z_{\rm eff}\approx 3Z$, which will further
lower the age estimate from eq.~(\ref{Age-L}) by about 5$\%$.

$Y_{\rm HB}-Y$ is estimated to be 0.01--0.02 in models without helium
diffusion \cite{IbenRenzini1984}, and $-$0.005--0.005 in helium diffusion
models \cite{ProffittMichaud1991,ProffittVandenberg1991,thiswork}.
Therefore a strong dependence of the age on the helium abundance $Y$ is
manifest in both eq.~(\ref{Age-LIben}) and eq. (\ref{Age-L}).

It is conventionally assumed that $Y\approx 0.24$ in GCs
\cite{Sandage93a,Bergbusch1992}, in agreement with the observed
primordial helium abundance $Y=0.235\pm 0.01$ \cite{Skillman1993}
and the prediction of Big Bang Nucleosynthesis \cite{Walker1991,Copi}.
But both eq.~(\ref{Age-LIben}) and eq. (\ref{Age-L}) show that
a higher $Y$ (but $\la$ 0.28) and can significantly reduce the age estimate
of GCs. For example, a $Y$ of 0.28 will reduce the age by as much as 20$\%$
with respect to a $Y$ of 0.24, according to eq.~(\ref{Age-LIben}) and
(\ref{Age-L}). Even a $Y$ of 0.26 will lower the current age estimate by
10$\%$.

Determining the age of GCs with the isochrone fitting method has a similar
$Y$ sensitivity as in eq.~(\ref{Age-LIben}) or (\ref{Age-L})
if a distance modulus from ZAHB fitting is used.
If a distance from ZAMS fitting is used in the isochrone fitting,
the effect of a higher Y in GCs on their age is not easy to quantify.
Figure 2 shows the comparison of a $Y=0.28$ isochrone and $Y=0.24$ isochrones
in fitting the fiducial sequence of NGC6397 (adopted from \cite
{Buonanno89}). Both isochrones include helium diffusion but don't include
oxygen-enhancement due to the adoption of Livermore opacities in most of
the temperature range, and have a $Z=0.0002$.
NGC6397 is believed to have [Fe/H]=-1.91 \cite{Sandage93a}.
The reddening and distance modulus adopted for the cluster are 0.11 and
12.35 for the $Y=0.24$ isochrones, 0.14 and 12.45 for the $Y=0.28$ isochrone.
The distance modulus estimated from calibrations of RR Lyrae stars
is about 12.5 \cite{Buonanno89}.
Fig. 2 shows that while NGC6397 has an age of $\sim 15$ Gyr when
fit with the $Y=0.24$ isochrones, it is only $\sim 12$ Gyr old when fit
with the $Y=0.28$ isochrones. This is in general agreement with
eq.~(\ref{Age-L}) from which a $Y$ of 0.28
yields an age 20$\%$ younger than a $Y=0.24$.

\section{The measurement of Helium abundances in GCs \label{HeliumObs.}}
The surface temperatures of stars in globular clusters are too cool to
have detectable helium lines. Therefore, the
$Y$ in GCs cannot be measured directly from spectra of stars.
Instead, $Y$ is indirectly determined from comparisons between certain
theoretically $Y$ dependent parameters of GCs deduced from stellar models
and the measurement of such parameters from GC observations.

One such parameter is the relation between the effective temperature
and period of RR Lyrae stars at the blue edge of the instability
strip \cite{TuggleIben1972}. Comparison of this relation between canonical
stellar models (i.e., no rotation, no diffusion, no mass loss, etc) and
observations yields $Y_{\rm HB}=0.27\pm 0.04$ \cite{Bingham1984}.
A similar comparison is done to the $A$ parameter, which
relates the effective temperatures and periods of general RR Lyrae stars
\cite{Caputo83}. This comparison usually yields a high
$Y_{\rm HB}$ of 0.26--0.33 \cite{Caputo83}, and leads to an anti-correlation
between the metallicity of a cluster and its helium abundance, which can hardly
be explained with any chemical evolution theories \cite{Caputo83,Sandage81}.
This anti-correlation problem may vanish if one assumes
a varying core mass in HB stars due to non-canonical effects
(for example, rotation \cite{Renzini77}), and/or
a varying CNO abundance \cite{Caputo83}.
Both of these comparisons, however, involve the determination of the effective
temperatures of RR Lyrae stars, which, as mentioned above,
cannot be confidently inferred from their observed colors due to
the uncertainty in the interstellar reddening and stellar modelling.

Another parameter is $R$, which is the ratio of the number of stars on the
horizontal branch to the number of stars on the red giant branch with
luminosities higher than the mean luminosity of RR Lyrae stars. Stellar models
show that $R$ is very sensitive to $Y_{\rm HB}$ and thus $Y$. Canonical
stellar models yield a $R$-$Y$ relation \cite{Buzzoni1983,CMP1987}:
\begin{equation}
Y=0.186+0.37\log R,
\label{Y-R}
\end{equation}
assuming $Y_{\rm HB}-Y\approx 0.02$, a result from canonical models that
exclude helium diffusion. Eq.~(\ref{Y-R}) yields an average
$Y$ of $0.23\pm 0.02$ for 15 GCs \cite{Buzzoni1983}.
However, if other non-canonical effects such as rotation operate in GC
stars to increase $M_{\rm c}$, the helium core mass of HB stars,
estimate of $Y$ can be higher than canonical
values. For example, a core rotation rate of 2$\times 10^{-4}$ rad/s, a
typical main sequence value, will increase $M_{\rm c}$
by 0.03$M_\odot$\cite{Bingham1984}, and increase the estimate of $Y$ from
the $R$ comparison by 0.01, since \cite{CMP1987}
\begin{equation}
dY/dM_{\rm c}=0.4.
\label{dY/dM}
\end{equation}
Assuming a larger and varying $M_{\rm c}$, Caputo {\sl et al.} obtained an
average $Y$ of $0.24\pm 0.01$ when combining the $R$-method and the
$A$-method \cite{CMP1987}. Furthermore, as we have shown, if we include
helium diffusion, $Y_{\rm HB}-Y$ will be only about 0.005. In addition,
$M_{\rm c}$ will be slightly increased by 0.005$M_\odot$ with respect to that
of canonical models due to the helium settlement into the core, which
results in a 0.002 increase in $Y_{\rm HB}$.
The resultant $Y$ will therefore be 0.02 higher than the value inferred from
eq.~(\ref{Y-R}), just from the inclusion of helium diffusion in stellar
models alone.

The luminosity spread of the horizontal branch serves as another parameter
that can be used to determine the helium abundance.
A larger $Y_{\rm HB}$ yields a broader luminosity spread among the
horizontal branch stars, according to HB models \cite{Sweigart}.
Such comparisons for 47 Tuc and M15 suggested a $Y_{\rm HB}$ close to
0.24 \cite{DormanVandenbergLaskarides1989,DormanLeeVandenberg91}.
But again, possible systematic variations in $M_{\rm c}$ unaccounted for
in canonical models may affect the result.
For example, to bring the $Y$ values from the $A$-method and the $R$-method
into agreement, it is shown by Caputo {\sl et al.}
that $M_{\rm c}$ in M15 has to be increased by 0.04$M_\odot$
relative to its canonical value \cite{CMP1987}.
According to the calculations of Sweigart and Gross \cite{Sweigart},
a 0.04$M_\odot$ increase in $M_{\rm c}$ will narrow the predicted luminosity
spread of a horizontal branch appreciably and thus allow a higher $Y_{\rm HB}$
given an observed luminosity width of HB.

A $\Delta$ paramter is also proposed to determine $Y$,
where $\Delta$ is defined as the magnitude difference
between RR Lyrae stars and the main sequence at $(B-V)_0$ (the corrected
$B-V$ color)=0.7 \cite{Caputo81}.
Comparison with the measured $\Delta$ yields a $Y$ of 0.2 to 0.3.
This method depends on the determination of the metallicity of GCs,
the color determination \cite{Caputo81}, and hasn't accounted for
effects such as helium diffusion.

Despite these many methods of determining $Y$ for GCs, a statistically
sound upper limit on $Y$ is lacking. However,
a $Y>0.28$ is very hard to reconcile with most of the
above determinations, especially the comparisons of the $R$ parameter
and the luminosity spread of HB,
even when aforementioned non-canonical effects are taken into account.

A maximal helium abundance of 0.28 in GCs seems also to be consistent with that
inferred from luminosities of the HB stars. $Y_{\rm HB}$ is related
with the luminosity of RR Lyrae stars by eq.~(\ref{HBLum}) in canonical models.
Current calibrations of the visual magnitude of RR Lyrae stars yield a 0.2--0.3
magnitude spread for different methods, and an uncertainty of
0.2 magnitude within each method
\cite{Walker1992,Sandage1993b,Carney1992,SandageCacciari1990}.
which translates into a 0.06 uncertainty in $Y_{\rm HB}$.
In other words, a $Y$ of 0.28 can be easily accomodated for individual
clusters, given a primordial value of 0.24.
It should be noted that the observation of field RR Lyrae stars
in the disk of the galaxy may not serve a valid constraint on
$Y_{\rm HB}$ in globular clusters if only stars in GCs are enriched in helium.
For example, RR Lyrae stars in seven Large Magellanic Cloud clusters
are on average 0.28 magnitude brighter than field RR Lyrae stars
\cite{Walker1992}, which can be consistent with an enhancement of
helium abundances only in globular clusters.

\section{Scenarios to enhance the helium abundance in GCs\label{Scenario}}
A helium abundance higher than the primordial value $Y_{\rm p}$
in GCs is possible to achieve under certain scenarios.
The simplest scenario is the enrichment by the first generation
supernovae (SNe) \cite{ShiSchrammDearbornTruran,BrownBurkertTruran1991}
Measurements of low metallicity HII regions in
irregular galaxies and extragalactic HII regions showed that
their helium abundance $Y$ correlates
with their oxygen abundance [O/H] by \cite{Skillman1993,Pagel1992}
\begin{equation}
Y\approx Y_{\rm p}+130{\rm [O/H]}\quad {\rm for}\quad
{\rm [O/H]}\la 2.5\times 10^{-4},
\label{Y-O}
\end{equation}
i.e., $dY/dZ\approx 6$. For GCs with [Fe/H]=$-1$, their oxygen abundances
are about 1/3 of the solar value \cite{Wheeler89},
or $[O/H]\approx 3\times 10^{-4}$. If GCs follow
the same helium-oxygen correlation as those low metallicity HII regions,
they have $Y\approx 0.27$, which is significantly above the primordial value.
Of course, it is debatable whether GCs follow the same $Y$-[O/H]
relation as low metallicity HII regions,
although both of them have similarly low metallicities.
The sun, on the other hand, has $Y\approx 0.28$ and
[O/H]$=8\times 10^{-4}$, which would require a different relation
$Y\approx Y_{\rm p}+60$[O/H] (i.e., $dY/dZ=3$).
Therefore, solar-type disk stars must have had quite a different chemical
history from low metallicity gases, like those of GCs and HII regions.
It is worth noting that VandenBerg, Bolte and
Stetson \cite{VandenbergBolteStetson1990} found
a 2--3 Gyr age spread among GCs with [Fe/H]$>-1.6$ and no significant age
spread among GCs with [Fe/H]$<-1.6$. If an enhancement in the helium abundance
correlates with an increase of metallicity in GCs, as suggested above,
the age spread among metal-rich GCs may be explained at least in part by
the different enrichment of helium in different clusters.

Exotic models, such as Pop III stars
\cite{CarrBondArnett1984,BondArnettCarr1984},
may enrich helium but not heavier elements. It is calculated that
if the oxygen cores of these Pop III stars are larger than $\sim 100M_\odot$,
the cores will collapse completely into blackholes
without ejecting heavy elements. However, processed helium in Pop III
stars can be deposited into the
interstellar medium through mass loss and mixing of Pop III stars.
The maximal amount of helium enrichment is $\Delta Y=0.17$ for an initial
helium abundance of 0.24, if all gas is processed once in Pop III stars
\cite{CarrBondArnett1984,BondArnettCarr1984}.
Therefore, in order to enhance the helium abundance in a $10^6M_\odot$
cloud from 0.24 to about 0.28, at least $2.5\times
10^5M_\odot$ gas has to be processed through Pop III stars.
This requires quite a different mass function from the
$M^\alpha$ power law with $\alpha=-2.35$
observed today \cite{Salpeter1955}. Since the metallicity of the
cloud is low, the required mass function could have $\alpha\ga -0.5$, so that
there were no more than $\sim 100$ SNe in a $10^6M_\odot$ cloud that
would deposit metals into the cloud.

An enhancement of the helium abundance may also arise during the
formation of globular clusters due to
the difference in the first ionization energy between the
hydrogen and the helium (13.6eV vs. 24.6eV).
In certain GC formation scenarios, this difference may enable
helium to recombine while hydrogen are still heavily ionized.
As a result, helium may have a slower sound speed than hydrogen and
is therefore more prone to gravitational collapse.

One class of models of forming GCs,
first proposed by Peebles and Dicke \cite{PeeblesDicke1968},
argues that GCs form from primordial
baryon density fluctuations after recombination, which have a Jeans mass
of a typical GC mass, 10$^6 M_\odot$.
According to the standard recombination picture of the universe, hydrogen
recombined at $T\sim 3000$K (redshift $z\sim 1000$) \cite{Peebles1993}
while helium recombined at $T\sim 10^4$K
or $z\sim 3000$ (inferred from Fig. 5 of ref. \cite{GouldThakur1970}).
In between $z\sim 1000$ and 3000 there existed a period when helium
recombined and primarily interacted with atoms, ions and electrons,
and hydrogen were still mostly ionized and primarily interacted with
background photons. Therefore helium can collapse earlier than hydrogen
due to its lower sound speed.

In the period after helium recombination and before hydrogen recombination,
the helium falls quickly into the potential well of dark matter,
following the density fluctuation of dark matter \cite{KolbTurner1990}.
As hydrogen recombines,
the density fluctuations in the hydrogen and the helium begin to grow
jointly. The overdense region, however, already has an enhanced helium
abundance from the in-fall of helium early on. Assuming the density fluctuation
in dark matter at $z\sim 1000$ to be $\delta\rho/\rho$ and the primordial
helium abundance $Y_{\rm p}=0.24$, the enhanced helium abundance
in the overdense region at $z\la 1000$ is \cite{ShiSchrammDearbornTruran}
\begin{equation}
Y={Y_{\rm p}(1+\delta\rho/\rho)\over Y_{\rm p}(1+\delta\rho/\rho)
+1-Y_{\rm p}}\approx 0.24(1+0.76\delta\rho/\rho).
\label{Peebles}
\end{equation}
To enhance the helium abundance to $Y=0.26$ on the 10$^6M_\odot$ scale
requires $\delta\rho/\rho\sim 0.1$ on this scale,
which means $\delta\rho/\rho$ on the 10$^6M_\odot$ scale
will become nonlinear at $z\sim 100$. This cannot be realized
in the standard cold dark matter (CDM) model, in which structures
on this scale become non-linear
at $z\la {\cal O} (10)$ \cite{KolbTurner1990}, but may be possible
in models where primordial seeds such as topological defects
provide nonlinear fluctuations on small scales
\cite{Vilenkin1985,TurokSpergel1991,LuoSchramm1993}.
A $\delta\rho/\rho$ of 0.1 at $z\sim 1000$ with $\sim$10$^6M_\odot$ scales
would correspond to fluctuations in the microwave background radiation
on scales that are less than 8 arc minutes and thus would be smoothed by the
finite size of the surface of last scattering.

In models that generically form $\sim 10^6M_\odot$ cold clouds in
proto-galaxies \cite{Fall1985,MathewsSchramm1993},
a higher helium content may arise if Pop II stars form in the shock
waves from the first generation supernovae in the cloud (simplified
in Figure 3) \cite{BrownBurkertTruran1993}. The temperature
of these clouds can stay around 10$^4$K for a long time
due to a sharp drop in the cooling
function of the gas cloud at 10$^4$K \cite{GouldThakur1970}.
If a strong UV photon background exists (which is likely due to the
presence of supernovae shocks),
hydrogen can be partially ionized at this temperature
and interact primarily with the UV photons in the cloud,
but helium remains mostly neutral and interacts primarily with
atoms, ions and electrons.
The sound speed in helium, $c_{\rm He}$, is then $\sim 10$ km/sec (for
a temperature of $\sim 10^4$K), while the sound speed in hydrogen,
$c_{\rm H}$, can be much larger. When the velocity of a supernova
shock wave drops below $c_{\rm H}$, the shock no longer compresses
hydrogen in the cloud into the shock front but continues to do so with
helium. The enhanced helium abundance is then \cite{ShiSchrammDearbornTruran}
\begin{equation}
Y\approx{Y_{\rm p} M_{\rm sh}\over (1-Y_{\rm p}) M_{\rm tot}\delta/l+Y_{\rm p}
M_{\rm sh}},
\label{SNshock}
\end{equation}
where $M_{\rm sh}$ is the mass swept by the shock, $M_{\rm tot}$ is the total
mass of the cloud, $Y_{\rm p}$ is the helium abundance before
the shock, $\delta$ is the thickness of the shock front and $l$ is
the thickness of the cloud after being compressed by the shock wave.
If the SNe rate is $\ga 10^{-6}$ yr$^{-1}$ per cloud, the shock wave can
propagate through the entire cloud at a speed larger than $c_{\rm He}$
\cite{ShiSchrammDearbornTruran,BrownBurkertTruran1991}.
Then $M_{\rm sh}\sim M_{\rm tot}$. If $l\gg \delta$, i.e., the density
inside the shell is much higher than the density outside,
$Y$ can be significantly larger than $Y_{\rm p}$.
Therefore stars formed from the compressed gas at the shock front may
have a higher helium abundance than the primordial value.

We have briefly discussed several scenarios to enhance the helium abundance
in GCs. There may be other possible scenarios, such as a magnetic field, that
may also segregate ionized gas from neutral gas.
If these scenarios operate primarily in GCs, they will not have a direct
effect on HII regions where the primordial helium abundance is measured,
nor will they affect the helium abundances in galactic disks.
Helium-enriched stars from disrupted GCs may only constitute a small
population of halo stars. Therefore, there may not be evidence in disk
stars that support helium enhancement processes.
Only more refined stellar evolution models
and GC observations can better determine the helium abundance in GCs.
Once again, a spread in $Y$ after these helium enhancement processes may
be a candidate to account at least in part for the spread in age
estimates for different GCs which assume a universal $Y$.

\section{The Uncertainty in Mass Loss \label{massloss}}
Possible mass loss near the turn-off region of GCs is another uncertainty
that affects the age estimate of GCs
\cite{ShiSchrammDearbornTruran,DearbornSchramm1993,WilsonBowenStruck-Marcell1987}.
It is motivated by the closeness of GC turn-off temperatures
and the small temperature range 6600$\pm 200$K, where Pop I F-type stars
are observed to have severely depleted lithium abundances (the so-called
lithium dip) \cite{BoesgaardTippico,HobbsPilachowski}.
One explanation of the lithium dip is an instability strip induced mass loss
\cite{SchrammSteigmanDearborn1990}.

Lithium burns rapidly at temperatures
above 2.5$\times 10^6$K. It is therefore completely destroyed at layers
with temperatures above 2.5$\times 10^6$K \cite{LarryBrown}.
At the temperature range of F stars in
which the lithium dip occurs, the base of a star's outer convective zone cannot
reach this lithium-burning temperature. Therefore, its
surface lithium abundance should be its primordial value (although some
negligible depletion occurs in the pre-main sequence \cite{Deliyannis90}).
The mass loss at the surface, however, leads to a deeper penetration of
the convective layer into the star's radiative interior.
When the convective zone reaches down to layers with no lithium,
the surface lithium abundance can be depleted due to
mixing \cite{SchrammSteigmanDearborn1990,SwensonFaulkner1992}.
The shape of the observed lithium dip on the [Li/H]-Temperature plane can be
reproduced quite well in the mass loss mechanism with a mass loss
rate \cite{SchrammSteigmanDearborn1990}
\begin{equation}
\dot M\ga 7\times 10^{-11}M_\odot {\rm yr}^{-1}.
\label{MLR1}
\end{equation}
But the narrowness of the observed Pop I lithium dip constrains any significant
mass loss to a restricted temperature range between 6500K and 6700K.
On the other hand, the absence of a severe beryllium depletion in this
temperature range in Pop I stars limits any mass loss rate
to \cite{SchrammSteigmanDearborn1990}
\begin{equation}
\dot M\la 1\times 10^{-10}M_\odot {\rm yr}^{-1},
\label{MLR2}
\end{equation}
which coincides with the limit from observations on
ionized stellar winds of F type main sequence stars \cite{Brown}.
Alternative explanations of the lithium dip include
radiative diffusion \cite{Michaud86}, and rotation-induced
mixings that leads to a more extensive of mixing between the surface
lithium and the depleted lithium in the radiative interior \cite{Charbonnel}.

The same instability strip which may induce mass loss in Pop I F-type stars
may also operate in Pop II stars
in a similar temperature range, 6500--6700K. These stars generally have
an undepleted lithium layer of about 0.04$M_\odot$ at ZAMS.
Given a maximal lifetime of several Gyr they spend in the temperature range,
Dearborn {\sl et al.} have shown that if the mass loss rates are
$\ga 10^{-11}M_\odot$/yr, it should produce a similar lithium dip
in Pop II stars \cite{DearbornSchrammHobbs}.
Several halo stars in the vicinity of this temperature range have been observed
to have depleted lithium, but the existence of a lithium dip in Pop II
stars is still far from established \cite{Thorburn}. It is to be noted,
however,
that even an absence of the lithium dip in Pop II stars doesn't automatically
exclude the instability strip induced mass loss,
as long as the mass loss rates are $\la 10^{-11}M_\odot$/yr.
This instability strip induced mass loss does not affect the
lower main sequence and so avoids the arguments against large-scale
mass loss through out the entire lower main
sequence \cite{SchrammSteigmanDearborn1990}.

The position of the instability strip on the Pop II main sequence is quite
uncertain. If mass loss operates in stars in the Pop I lithium dip,
by a simple extrapolation, mass loss may also operate in a similar
temperature range around 6500--6700 K, which is in
the vicinity of the turn-off temperature of GCs \cite{DearbornSchramm1993}.
Observations of low metallicity blue stragglers, on the other hand,
seems to indicate that the instability strip is
bluer than the turn-off point of GCs with similar metallicities
\cite{Carney1994}. However, blue stragglers may be too pathological to
draw conclusions on the main sequence of GCs.

Another justification for instability strip induced mass losses in the
turn-off region of GCs may come from the red edge of
RR Lyrae stars and the calculated dependence of this red edge on
$Y$ (the envelope helium abundance), $Z$ (the metallicity),
$M$ (the mass) and $L$ (the luminosity).
Observations of RR Lyrae stars found the red edge of the instability
strip in HB stars lies at $\log T_{\rm e}\sim$3.81 for M15 and
$\log T_{\rm e}\sim$3.79 for M3. Although there hasn't been a theoretical
calculation of the position of the instability strip for Pop II stars,
it has been shown for Pop I cepheids \cite{ChiosiWoodBertelli}
\vfill\eject
\begin{eqnarray}
\log T_{\rm FRE}=&&({\rm const.}+0.106Y)+(0.163-0.392Y-0.592Z)\log (M/M_\odot)
\nonumber\\
&&+(-0.074+0.074Y-0.437Z)\log (L/L_\odot),
\label{Cepheids}
\end{eqnarray}
where $T_{\rm FRE}$ is the effective temperature of the foundamental red edge.
If we take eq.~(\ref{Cepheids}),
and take $Y_{\rm HB}\approx 0.25$, $Z<10^{-3}$,
$M/M_\odot\approx 0.65$ for RR Lyrae stars,
$Y\la 0.1$ \cite{ProffittVandenberg1991,thiswork},
$M/M_\odot\approx 0.75$ for main sequence turn-off stars, and
$L_{\rm RR}/L_{\rm TO}\approx 1.4$ \cite{Sandage93a},
we get $\log T_{\rm FRE}(\rm RR)-\log T_{\rm FRE}(TO)\approx -0.05$.
Therefore, the instability strip at the main sequence turn-off luminosity
has a red edge of $\log T_{\rm FRE}\sim 3.85$, which is very close to the
turn-off region of metal-poor globular clusters ($T_{\rm e}\sim 3.80$ to 3.83
\cite{ProffittVandenberg1991,thiswork}). It is therefore conceivable
that a consistent instability strip calculation for Pop II stars can extend the
instability strip to the turn-off region. In this regards,
theoretical calculations of the instability strip for Pop II stars are
necessary to extrapolate the red
edge accurately from RR Lyrae stars down to the turn-off luminosity.

When mass loss occurs at the temperature of the turn-off, the
small reduction in mass causes the model to turn off at a lower luminosity
with respect to models without mass losses, since the turn-off luminosity of
a main sequence Pop II star decreases as the mass of the star
decreases \cite{DearbornSchramm1993,WilsonBowenStruck-Marcell1987}.
Stars also appear to spend an increased fraction of their life beyond the
turn-off (since less massive stars always evolve slower than more massive ones)
\cite{DearbornSchramm1993}.
The GCs will then (incorrectly)
appear older due to their lower turn-off luminosities,
according to eqs.~(\ref{Age-LIben}), (\ref{Age-LSandage}) and (\ref{Age-L})
\cite{DearbornSchramm1993,WilsonBowenStruck-Marcell1987}.

Several authors \cite{ShiSchrammDearbornTruran,DearbornSchramm1993}
have calculated the effect of a mass loss rate
$\sim 10^{-11}M_\odot$ yr$^{-1}$ in the temperature range of $6500\pm 200$K
and $6600\pm 200$ K on the evolution of GCs. Such assumptions result in GCs of
11 Gyr to 13 Gyr old looking 2$\sim 3$ Gyr older.
To see the effect of the mass loss on our helium diffusion models,
we repeated our model calculations in section 2 with a mass loss rate
\begin{equation}
{\dot M}=10^{-11}M_\odot {\rm yr}^{-1}\exp\Bigl[-({T-6500{\rm K}
\over 240{\rm K}})^2\Bigr],
\label{MLR3}
\end{equation}
for both $Y=0.24$ and $Y=0.28$. Figure 4 shows the resultant
turn-off visual magnitude vs. age relation for different cases.
We find that a mass loss rate of eq.~(\ref{MLR3})
will only affect isochrones younger than $\sim 12$ Gyr for $Y=0.24$ and
$\sim 15$ Gyr for $Y=0.28$ significantly,
decreasing their turn-off visual magnitude to
the level of about 1 Gyr older isochrones without mass loss.
Therefore, for a GC age of 14--15 Gyr by conventional estimates, or
about 11--12 Gyr if assuming $Y=0.28$,
the decrease in age due to a mass loss rate of eq.~(\ref{MLR3})
alone is at most $\sim 1$ Gyr. This age reduction due to mass loss
is smaller than the 2--3 Gyr reduction
calculated previously from stellar models without helium diffusion
\cite{ShiSchrammDearbornTruran,DearbornSchramm1993}.
The reason may lie in that in helium diffusion models,
mass loss can strip the outer helium-poor layer of stars and thus reduce the
effect of helium diffusion on stellar evolution. For example,
a 0.8$M_\odot$ low-$Z$ model with no mass loss has a surface helium abundance
of less than $3\%$ at turn-off due to helium diffusion, but a similar model
with a mass loss of eq.~(\ref{MLR3}) has a surface helium abundance of
more than $20\%$.

Since it has been shown that a high helium abundance of 0.28 can decrease
the age estimate for GCs from $\sim 14$ Gyr to about 11 Gyr,
from figure 4 it can be seen that a combination of $Y=0.28$ and a mass loss
rate like that in eq.~(\ref{MLR3}) can further decrease the GC age to as low as
10 Gyr. However, to what extent the
mass loss mechanism can lower the age estimate of GCs is not only determined
from its turn-off magnitude-age relation but also constrained by
the calculated GC luminosity function under mass loss compared with
observations
\cite{DearbornSchramm1993,SwensonFaulkner1992}.

It has been pointed out that mass loss near the GC turn-off point can
significantly affect the luminosity function of GCs,
since the longer life time that stars spend after their turn-off due to mass
loss results in a bump on the LF in the
subgiant region \cite{DearbornSchramm1993,SwensonFaulkner1992}.
In figure 5(a) and 5(b), we plot the predicted luminosity functions of
GCs with a mass loss rate of eq.~(\ref{MLR3})
(assuming a Salpeter initial mass function \cite{Salpeter1955})
for both $Y=0.24$ and $Y=0.28$, compared with observations compiled by
Stetson \cite{Stetson}. For $Y=0.24$, figure 5(a) shows that
a 12 Gyr mass loss LF can be
ruled out by the absence of any large size bump
at the subgiant region in the observation, but a 13 Gyr mass loss
LF may still be allowed. For the $Y=0.28$ case,
figure 5(b) shows that the 11 Gyr
mass loss LF can be confidently ruled out, but the 12 Gyr mass loss LF may not
be ruled out easily.  Therefore, due to the constraint from
the luminosity function of GCs, mass loss essentially
cannot lower the age of GCs to below 12 Gyr even when combined with
a high helium abundance.
This result should also hold if we consider
oxygen-enhancement in GCs, or if we use another power-law
initial mass function, since it has been shown that the LFs of GCs in
the turn-off and subgiant region are not sensitive to both of the two factors
\cite{Stetson}.

It is interesting to note from figure 5(a) and (b) that
mass loss can improve the fit at the red giant region of LF between
models and observations. But it may be premature to draw any conclusion from
comparison to this uncertain portion of the data.

Isochrone fittings also seem to indicate that even with a combination of
high $Y$ and mass loss, an age lower than 12 Gyr is hard to reach.
Figure 6 shows $Z=0.0002$ isochrones with eq.~(\ref{MLR3}) and $Y=0.28$ at
12 and 13 Gyr old, overlaid on the fiducial sequence of NGC6397,
as well as a standard isochrone without mass loss and with
$Y=0.24$ at 15 Gyr (the same one as in Fig. 2). Clearly,
an isochrone in between 12 Gyr old and 13 Gyr old
with $Y=0.28$ and mass loss may fit the observation
acceptably (while the 12 Gyr isochrone with $Y=0.28$ and no mass loss
fits fairly well in figure 2). Therefore, effects of a higher $Y$
and a mass loss on GC ages do not add up linearly in the isochrone
fitting method, and an GC age lower than 12 Gyr seems still unachievable.

Applying our mass loss calculations to Pop II stars, we find that
with a peak mass loss rate of
$10^{-11}M_\odot$ yr$^{-1}$, severe lithium depletions occur at stars
between $\sim 0.8M_\odot$ and $\sim 1.0M_\odot$ (similar to the conclusion
of Dearborn {\sl et al.} \cite{DearbornSchrammHobbs}). Therefore, in
isochrones older than $\sim 14$ Gyr for $Y=0.24$ or older than $\sim 12$ Gyr
for
$Y=0.28$, no stars with lithium severely depleted from mass loss will be
observed since they already evolved away from the red giant branch. That
is, if further observations find no evidence for a mass loss induced
lithium depleted in Pop II stars, it may simply indicate that our universe is
too old for those lithium depleted Pop II stars (if there were any) to survive.
In isochrones younger than these ages, lithium-depleted stars should be
observed
in either their main sequence, subgiant or giant phase. It should be noted
that a difference in mass loss rates may exist between globular clusters and
field stars if only the helium abundance in globular clusters is enhanced,
since a higher helium abundance increases the surface temperature of
stars and therefore shifts the relative position of the instability strip in
isochrones.

\section{Summary\label{Summary}}
As we have seen, the possible systematic uncertainties in the helium
abundance of globular clusters and an instability strip induced mass loss
near the turn-off region of globular clusters allow an average age of globular
clusters as low as $\sim 20\%$ below current estimates.
Therefore the currently quoted age estimate of
14.1$\pm 1.5$ Gyr \cite{Sandage93a} or 15$\pm 1.5$ \cite{Walker1992}
is potentially subject to systematic shifts and its
uncertainties are model dependent and not fully represented by the quoted
error. Based on eq.~(\ref{Age-L}) and our mass loss calculations,
we recalculated the age of five metal-poor clusters using data compiled by
Sandage \cite{Sandage93a} for four cases with and without oxygen-enhancement.
Tabulated in table 1, the resultant average ages range from 11.3$\pm 1$ Gyr to
14.6$\pm 1.5$ Gyr for the four cases, compared with 14.1 Gyr from Sandage's
estimate assuming a primordial helium abundance, no mass loss, no helium
diffusion and oxygen-enhancement. In table 1, oxygen-enhancement is
approximated by models with solar-scaled abundances that satisfy
eq.~(\ref{Oxygen}) \cite{Salaris}.
In case (iv), the constraint from the luminosity function of globular clusters
on mass loss holds true with and without oxygen-enhancement,
and constrains the age of globular clusters to be older than
$\sim 12$ Gyr in both cases.

It is unlikely that all globular clusters
will have an enhanced $Y$ of 0.28, given the current inferences of
the helium abundance in globular clusters. An average $Y$ of 0.26,
for example, is more compatible with available helium inferences.
In all cases, however, after uncertainties in the helium abundance, mass
loss, helium diffusion and oxygen enhancement are considered, and
the $\sim 10\%$ uncertainty from calibrations of RR Lyrae stars is included,
the lowest age estimate for globular clusters
seems to be about 10 Gyr.

Clearly, the true accuracy of globular cluster
age estimates would benefit from better
estimates of non-canonical effects in stellar models and
better determinations of mass losses in Pop II stars.
In inferring the helium abundance,
stellar model comparisons with observed GC parameters should
include non-canonical effects such as helium diffusion, core rotation, etc.
The existence of an instability strip induced mass loss
near the turn-off region can be explored by the further observations
of the luminosity functions of globular clusters and lithium observations
in Pop II stars.

A globular cluster age of 11$\pm 1$ Gyr is entirely consistent with the
age of the beginning of nucleosynthesis of heavy elements (i.e., SNe events)
measured by radioactive dating that gives a lower limit of 10
Gyr \cite{MeyerSchramm1986,CowanThielemannTruran1991}.
It is also consistent with the age of the galactic disk, 10$\pm 2$ Gyr,
measured from the cooling of white dwarfs \cite{Winget1987},
provided that the disk collapsed from the halo within 1 to 2 Gyr.

If globular clusters are indeed 11$\pm 1$ Gyr old, a universe as young as
11 Gyr is then possible if globular clusters formed at redshift
$z\ga 4$. This will significantly relax the current constraints on
cosmological parameters. For a baryonic universe where $0.01\la\Omega\la 0.1$,
the Hubble constant can be as high
as 90 km/sec/Mpc; for non-baryonic dark matter universes with
$\Omega> 0.3$, $H\la 77$km/sec/Mpc;
if $\Omega=1$, $H=60$km/sec/Mpc. To turn the problem around,
if the measurements of the Hubble
constant eventually converge, it will put constraints on $\Omega$ in
a $\Lambda=0$ Friedman-Robertson-Walker universe and improve
our understanding of the age of globular clusters.
\section{Acknowledgements}
The author sincerely thanks David Schramm, James Truran and David Dearborn
for their help and support. The author also thanks Bruce Carney, Ben Dorman,
Brian Fields, Sangjin Lee, Grant Mathews, P. B. Stetson,
Michael Turner and Don York for
valuable discussions. This work is supported by the DoE (nuclear), by NASA
and by NSF grant through grant AST90-22629 at
the University of Chicago, and by the DoE and NASA through grant 2381 at
Fermilab.
\vfill\eject

\vfill\eject
\begin{table}
\caption{The age estimates from eq.~(5) and our
mass loss calculations for five metal-poor globular clusters. {\sl All ages
suffer an additional $\sim\pm 10\%$ uncertainty from the $\sim\pm 0.1$
magnitude uncertainty in the calibration of RR Lyrae stars.} \label{table1}}
\begin{tabular}{ccccc}
Clusters& Case (i)\tablenotemark[1]
& Case (ii)\tablenotemark[2] & Case (iii)\tablenotemark[3]
& Case (iv)\tablenotemark[4] \\\hline
[O/Fe]=0&&&&\\
\ \ M68\ &12.5 Gyr&11.3 Gyr&10.2 Gyr&\\
\ \ M92\ &16.2 Gyr&14.6 Gyr&13.2 Gyr&\\
\ NGC6397&15.5 Gyr&14.0 Gyr&12.7 Gyr&\\
\ \ M15\ &14.5 Gyr&13.1 Gyr&11.8 Gyr&\\
\ \ M30\ &14.3 Gyr&12.9 Gyr&11.7 Gyr&\\\hline
Average&&&&\\
$[$O/Fe$]$=0&14.6 Gyr&13.2 Gyr&11.9 Gyr&$\ga 12$ Gyr\\
$[$O/Fe$]$=0.6&13.9 Gyr&12.5 Gyr&11.3 Gyr&$\ga 12$ Gyr\\
\end{tabular}
\tablenotemark[1]{$Y=0.24$, no mass loss;}
\tablenotemark[2]{$Y=0.26$, no mass loss;}
\tablenotemark[3]{$Y=0.28$, no mass loss;}
\tablenotemark[4]{an enhanced $Y$ and an instability strip induced mass loss
according to eq.~(15).}
\end{table}
\vfill\eject
\centerline{\bf Figure Captions}
\noindent Figure 1. Two isochrones with similar parameters but
different mixing lengths.
\medskip

\noindent Figure 2. Isochrones with and without an enhancement of the
helium abundance in GCs compared with the fiducial sequence of NGC6397
([Fe/H]$\approx -2$) (denoted by triangles).
The $Y=0.24$ isochrones are shifted by $E(B-V)$=0.11 and $m-M=12.35$;
the $Y=0.28$ isochrone is shifted by $E(B-V)$=0.14 and $m-M=12.45$.
\medskip

\noindent Figure 3. The propagation of a SN shock front inside a 10$^6M_\odot$
cloud.
\medskip

\noindent
Figure 4. The calculated turn-off magnitude-age relation
for different isochrones. Comparisons can only be made between
curves with similar $Y$, since different $Y$'s yield
different ZAHB luminosities.
\medskip

\noindent
Figure 5. The calculated luminosity functions for GCs compared with
observations \cite{Stetson}. $N$ is the number of stars in each 0.2 magnitude
bin. The reference point is defined as the
point on main sequence that is 0.05 mag redder than the turn-off.
(a) $Y=0.24$; (b) $Y=0.28$.
Mass losses are calculated assuming a mass loss rate of eq.~(\ref{MLR3}).
The sharpness of the bump in subgiant regions is due to the finite grid
of stellar models.
\medskip

\noindent Figure 6. Isochrones with and without a higher helium abundance
and mass loss, compared with the fiducial sequence of NGC6397 ([Fe/H]
$\approx -2$) (denoted by triangles). The 15 Gyr isochrone with
$Y=0.24$ and no mass loss is shifted by $E(B-V)=0.11$ and $m-M$=12.35;
the 12 and 13 Gyr isochrones with $Y=0.28$ and a mass loss
rate that satisfies eq.~(\ref{MLR3}) are shifted by $E(B-V)=0.14$ and
$m-M=12.38$.

\end{document}